\documentclass{PoS}

\title{Neutrons from multifragmentation reactions}

\ShortTitle{Neutrons from multifragmentation reactions}

\author{\speaker{W. TRAUTMANN}\\%
        GSI Helmholtzzentrum Darmstadt, Germany\\
        E-mail: \email{w.trautmann@gsi.de}}
\author{A.S.~BOTVINA\\
        INR Moscow, Russia}
\author{J.~BRZYCHCZYK\\
	Jagiellonian University, Krak{\'o}w, Poland}
\author{N.~BUYUKCIZMECI\\
        Sel\c{c}uk University, Konya, Turkey}
\author{I.N.~MISHUSTIN\\
        FIAS, Goethe Universit\"{a}t, Frankfurt, Germany}
\author{P.~PAW{\L}OWSKI\\
        IFJ-PAN, Krak{\'o}w, Poland}
\author{ALADIN2000 COLLABORATION}

\abstract{The neutron emission in the fragmentation of stable and radioactive Sn and La 
projectiles of 600 MeV per nucleon has been 
studied with the Large Neutron Detector LAND coupled to the ALADIN forward spectrometer 
at SIS. 
A cluster-recognition algorithm is used to identify individual particles within the 
hit distributions registered with LAND.
The obtained momentum distributions are extrapolated over the full phase space occupied
by the neutrons from the projectile-spectator source. 
The mean multiplicities of spectator neutrons reach values of up to 12 and depend strongly
on the isotopic composition of the projectile. An effective source temperature of
$T \approx 3 - 4$~MeV is deduced from the transverse momentum distributions. 
\newline
For the interpretation of the data, calculations with the Statistical Multifragmentation 
Model for a properly chosen ensemble of excited sources were performed.
The possible modification of the liquid-drop parameters of the fragment description in the hot
environment is studied, and a significant reduction of the symmetry-term coefficient
is found necessary to simultaneously reproduce the neutron multiplicities and the 
mean neutron-to-proton ratios $<$$N$$>$$/Z$ of $Z \le 10$ fragments. 
Because of the similarity of the freeze-out conditions with those encountered in 
supernova scenarios, this is of astrophysical interest.
}

\FullConference{ XLIX International Winter Meeting on Nuclear Physics\\
		 24-28 January 2011\\
		 BORMIO, Italy}

\begin{document}

\section{Introduction}
Neutron emission is an efficient means of cooling excited nuclei produced in nuclear 
reactions. For this reason, e.g., (heavy ion,xn) reactions have been and still are 
a standard tool 
for the spectroscopy of nuclei at high spin. Heavy projectiles will transfer large amounts 
of energy and angular momentum to the formed compound nuclei. Neutron evaporation then 
removes most of the energy but very little of the angular momentum, so that high-spin 
states close to the yrast line will be fed by the subsequent gamma decay.
Superdeformed bands at the highest spins observed 
have been discovered in this way~\cite{nolan_twin88}. 
By emitting neutrons the product nuclei also change their isotopic composition,
permitting high-spin spectroscopy deep into the neutron-poor region of the chart of nuclides. 
Neutron emission is thus important in the energetic evolution of heavy-ion reactions and, 
at the same time, related to isotopic phenomena.

\begin{figure}[htb!]
\centering
\includegraphics*[width=120mm]{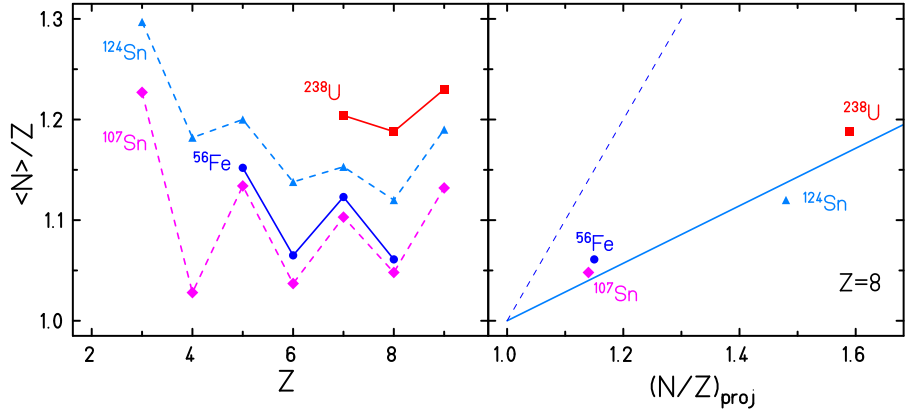}
\vskip -0.2cm
\caption{Inclusive mean values $<$$N$$>$/$Z$ of light fragments with $3 \le Z \le 9$ 
produced in the fragmentation of $^{107,124}$Sn (600 MeV/nucleon, ALADIN),
$^{56}$Fe and $^{238}$U (both 1 GeV/nucleon, FRS) as a function  
of the fragment $Z$ (left panel). The values for $Z=4$ have been corrected for the 
missing yield of unstable $^8$Be by including an estimate for it obtained from a smooth 
interpolation over the identified yields of $^{7,9-11}$Be. 
This correction 
makes the systematic odd-even variation more clearly visible for the neutron rich case. 
The right panel shows the results for $Z=8$ as a function of the $N/Z$ value of the 
projectile. The lines represent the trend of the data (full line) and 
$<$$N$$>$/$Z = (N/Z)_{\rm proj}$ (dashed); from Ref.~\protect\cite{traut07nn}.
}
\label{fig:noverz}  
\end{figure} 
Long evaporation cascades will include charged particle emissions which
will eventually compete with neutron emission in neutron-poor nuclei. 
This has led to the concept of a universal 
evaporation attractor line (or fragmentation corridor) for the location of final residues
in the plane of neutron number $N$ vs. atomic number $Z$. Predictions have been calculated with statistical models 
or were empirically deduced from experimental data~\cite{charity98,enqvist99,suemmerer00}. 
They are very useful for a wide range of reactions but deviations exist. 
In multifragmentation reactions, observed at excitation energies above a threshold of 
about 3 MeV per nucleon, the neutron content of produced fragments depends significantly
on the $N/Z$ ratio of the reaction system.
This is illustrated in Fig.~\ref{fig:noverz} with data measured with the ALADIN forward 
spectrometer and the FRS fragment separator at GSI~\cite{ricci04,napo04,sfienti09}. 

Multifragmentation is a rapid breakup of the highly excited reaction system and
associated with a copious production of nuclear fragments in addition to neutrons and 
light charged particles (for a review see, e.g., Ref.~\cite{dyntherm}). 
Statistical models have been found very useful for describing the partitioning of the 
excited systems~\cite{gross90,smm,botv95,MSU,Dag,EOS,INDRA,FASA}. 
They have also been used to explore possible modifications of fragment
properties in the hot environment of the freeze-out state. This is of interest because the
temperatures and lower than normal densities at freeze-out are similar to conditions 
encountered in supernova scenarios~\cite{ishi03,botv04}. 

As shown very recently for the case of multifragmentation of relativistic 
projectiles~\cite{ogul11},
the observed neutron richness of the produced intermediate-mass fragments requires a
significant reduction of the symmetry term in the liquid-drop description used for
them in the Statistical Multifragmentation Model (SMM, Ref.~\cite{smm}). 
With more neutrons bound in fragments, the multiplicity of free neutrons should be lowered,
thereby possibly offering an additional possibility for exploring fragment properties at 
freeze-out. 
This expectation can be tested since the Large Area Neutron Detector LAND~\cite{LAND} 
was used in these experiments to detect neutrons from the breakup of the projectile and
subsequent decays. Results from this study will be reported here.

\begin{figure}[htb!]
\centering
\includegraphics*[width=80mm]{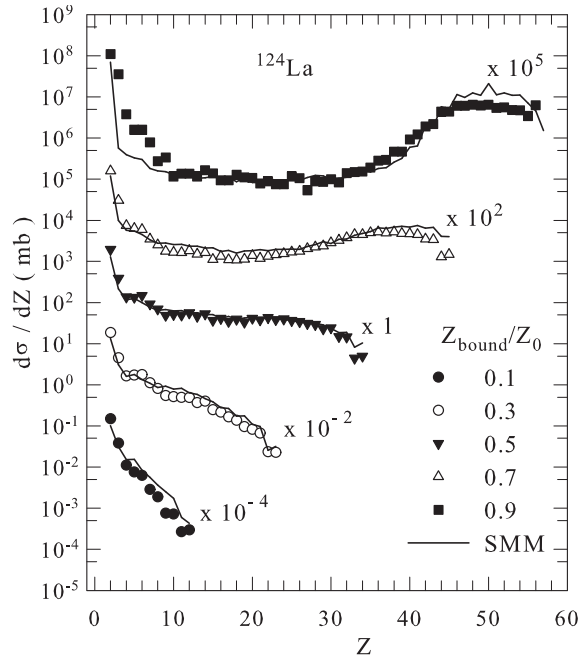}
\vskip -0.2cm
\caption{Experimental cross sections d$\sigma$/d$Z$ for the fragment production following
collisions of $^{124}$La projectiles sorted into five intervals of $Z_{\rm bound}/Z_0$ with centers
as indicated and width 0.2 (symbols) in comparison with normalized SMM calculations 
(lines, $Z_0 = Z_{\rm proj}$). 
Different scale factors were used for displaying the cross sections as indicated
(from Ref.~\protect\cite{ogul11}).
} 
\label{fig:ogul12}   
\end{figure}

In multifragmentation experiments, the detection of neutrons has also been essential for
investigating the transfer of energy. In previous ALADIN experiments, the excitation 
energy of the produced spectator systems was determined by using in
particular also the neutron energies and multiplicities measured with the LAND 
detector~\cite{poch95}. Multifragmentation experiments including the measurement of 
neutrons, although difficult because of the different techniques required for 
simultaneously detecting neutral and charged particles, have been performed at several 
laboratories~\cite{kunde96,sobot00,theri05,famiano06,wang07}. The problems and 
successes encountered in the calorimetry of multifragmenting systems have recently been 
reviewed~\cite{viola06}.

\section{Isospin dependent multifragmentation}

Isotopic effects in projectile multi-fragmentation at relativistic energies
were studied with experiment S254 of the ALADIN collaboration, 
conducted in 2003 at the SIS heavy-ion synchrotron~\cite{sfienti09,ogul11}. 
Besides stable $^{124}$Sn beams, neutron-poor secondary $^{107}$Sn and $^{124}$La beams
provided by the FRS fragment separator, all with 600 MeV/nucleon incident
energy, were used in order to explore a wide range of isotopic compositions. 
For the interpretation of the data, calculations with the Statistical Multifragmentation 
Model~\cite{smm} for a properly chosen ensemble of excited sources, adapted to the
participant-spectator scenario at relativistic energies, were performed.
The ensemble parameters were chosen so as to best
reproduce the $Z$ spectra and correlations of the observed fragments. 

\begin{figure}[htb!]
\centering
\includegraphics*[width=80mm]{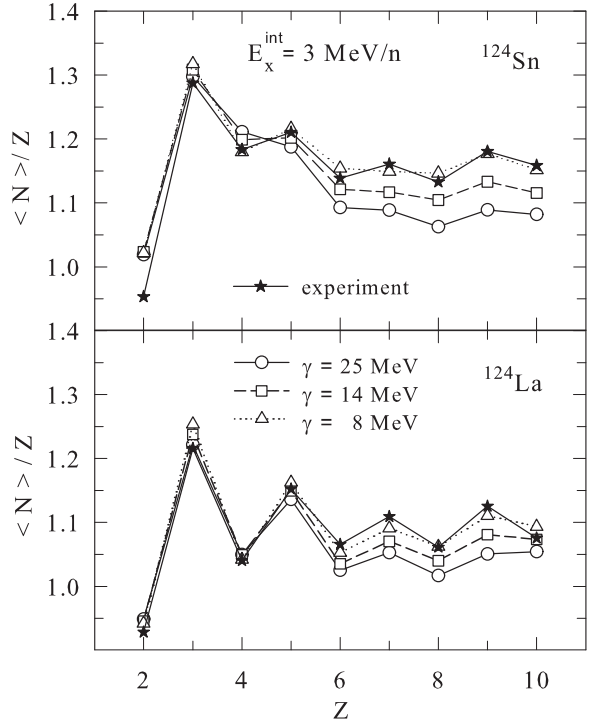}
\vskip -0.2cm
\caption{Mean neutron-to-proton ratio $\langle N \rangle /Z$ of fragments
produced in the fragmentation of $^{124}$Sn (top panel) and $^{124}$La (bottom panel) 
projectiles in the range $0.4 \le Z_{\rm bound}/Z_{\rm proj} < 0.6$. 
The experimental results (stars) are compared with SMM calculations using 
the three indicated values of the symmetry-term coefficient $\gamma$ and 
$E_x^{\rm int} = 3$~MeV/nucleon as the start value of the interpolation interval for the 
secondary-decay stage of the calculations. With this choice, $\gamma = 8$~MeV (triangles) 
gives the best agreement with the data.
} 
\label{fig:ogul22}   
\end{figure}

As an example of the good description achieved, the measured cross sections 
d$\sigma$/d$Z$ for fragment production in reactions initiated by $^{124}$La 
projectiles are shown in Fig.~\ref{fig:ogul12}. The data have been sorted into 
five bins of the
reduced bound charge $Z_{\rm bound}/Z_{\rm proj}$ of width 0.2, spanning the range up to 
$Z_{\rm bound} = Z_{\rm proj}$ ($Z_{\rm bound} = \Sigma Z_i$ for $Z_i \geq$~ 2). 
The charge distribution evolves from a so-called
'U-shaped' distribution, with domination of evaporation of the compound
nucleus and asymmetric binary decays, to a rapidly dropping exponential distribution. 
This evolution is a well-known characteristic feature and has been related
to the nuclear liquid-gas phase transition~\cite{ogilvie91,kreutz,hauger00}. 

The successful description of charge observables provides the basis for the 
investigation of the isotope distributions and the information contained therein.
As shown in more detail in Ref.~\cite{ogul11}, it is mostly the symmetry-term 
coefficient $\gamma$ which influences the isotopic composition of fragments. Model 
calculations performed for three values of $\gamma$~=~8, 14, and 25 MeV are shown 
in Fig.~\ref{fig:ogul22}. The mean neutron-to-proton ratio $\langle N \rangle /Z$
increases with decreasing $\gamma$, most noticeably for $Z \ge 7$. A reduced symmetry
term leads to larger widths of the initial isotope distributions at freeze-out and to
a reduced driving force toward the line of stability during subsequent decays. 
This effect is larger for the more neutron-rich system $^{124}$Sn.
At $Z \le 6$, the mean neutron-to-proton ratios exhibit the strong odd-even effects, 
typical for this type of reaction (Fig.~\ref{fig:noverz}).

While the fragment properties may be modified at freeze-out and during
the first deexcitation steps, as the hot fragments are still surrounded by other 
species, the standard properties must be restored at the end of the evaporation cascade.
In the calculations, a linear interpolation between these two limiting cases 
has been introduced in the interval of excitation energies between zero and a value
$E_x^{\rm int}$ which has been varied between 1 and 3~MeV/nucleon. 
Energy and momentum conservation are observed throughout this process. The results
shown in Fig.~\ref{fig:ogul22} were obtained for $E_x^{\rm int} = 3$~MeV/nucleon.
In this case, the best agreement with the experimental results is obtained with
$\gamma = 8$~MeV, while $\gamma = 14$~MeV appears to be the better choice if the 
interpolation towards the standard $\gamma = 25$~MeV is delayed until the 
excitation energy is at or below 1~MeV/nucleon~\cite{ogul11}.

\begin{figure} [tbh]
\centerline{\includegraphics[width=10cm]{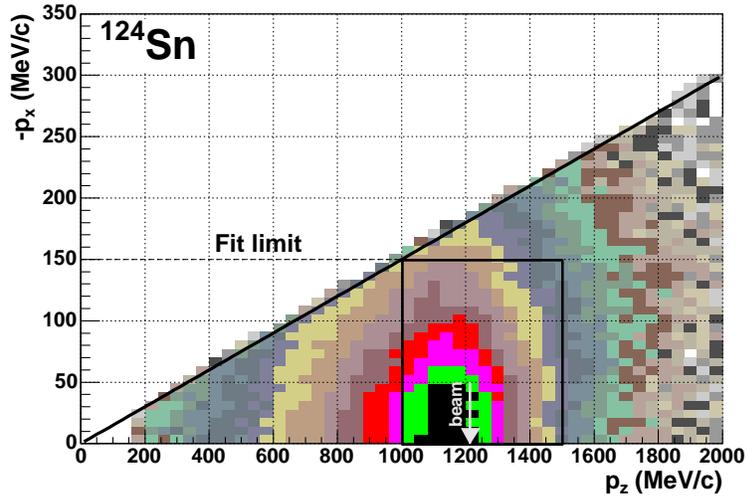}}
\caption{Distribution of identified neutrons in the plane of transverse momentum $p_x$
versus longitudinal momentum $p_z$ for the fragmentation of $^{124}$Sn projectiles
in collisions with natural Sn targets at an incident energy of 600 MeV/nucleon. The full
line indicates the acceptance cut in the analysis. The thin rectangle indicates the
two-dimensional interval $|p_x| \le 150$~MeV/c and $1000 \le p_z \le 1500$~MeV/c used
for the determination of the source temperature. 
}
\label{fig:source}
\end{figure} 

A similar result was obtained from the isoscaling analysis performed with the same data and 
discussed in Ref.~\cite{ogul11}. The isotopically resolved fragment yields provide 
evidence for a reduced symmetry energy in the hot environment at freeze-out, in accordance 
with previous findings~\cite{LeFevre,Botvina05}. 
Similar observations were made in other reaction studies performed at intermediate 
and relativistic energies~\cite{iglio06,souliotis07,hudan09,henzlova10}.

\section{Neutron detection and analysis}

The Large-Area-Neutron-Detector LAND is a 2x2x1 m$^3$ calorimeter consisting of in total 
200 slabs of interleaved iron and plastic strips viewed by photomultiplier tubes at both 
ends~\cite{LAND}. In the ALADIN experiments, the detector was positioned approximately
10 m downstream from the target and slightly asymmetrically with respect to the incoming beam
direction, so as to allow the deflected beam passing along it on one 
side~\cite{schuett96}. The angular acceptance of -9.8$^{\circ}$ to 1.2$^{\circ}$ in
horizontal and $\pm$5.6$^{\circ}$ in vertical directions
corresponded to a geometrical acceptance in perpendicular momenta $p_x$ from approximately
-210 MeV/c to 25~MeV/c and $|p_y| \le 120$~MeV/c for neutrons with a kinetic energy of 600 MeV. To avoid edge effects, the acceptance was reduced in the analysis.
A veto wall of 5-mm-thick plastic scintillators in front of the detector permitted 
the distinction of neutral and charged particles.

In the analysis, a cluster-recognition algorithm is used to identify individual particles 
within the registered hit distributions. The mean number of hits belonging to an
observed neutron event was found to be about 1.5. 
Particle momenta were determined from the neutron time-of-flight measured with a
resolution of $\Delta t \approx 550$~ps (FWHM).
For the case of $^{124}$Sn projectiles, the obtained distribution of neutron events in the 
plane of transverse momentum $p_x$ 
versus longitudinal momentum $p_z$ is shown in Fig.~\ref{fig:source}. 
The source of neutrons emitted from the projectile spectator is clearly visible, centered
near the projectile momentum $p_{\rm proj} = 1216$~MeV/c/nucleon. 

\begin{figure} [tbh]
\centerline{\includegraphics[width=8.5cm]{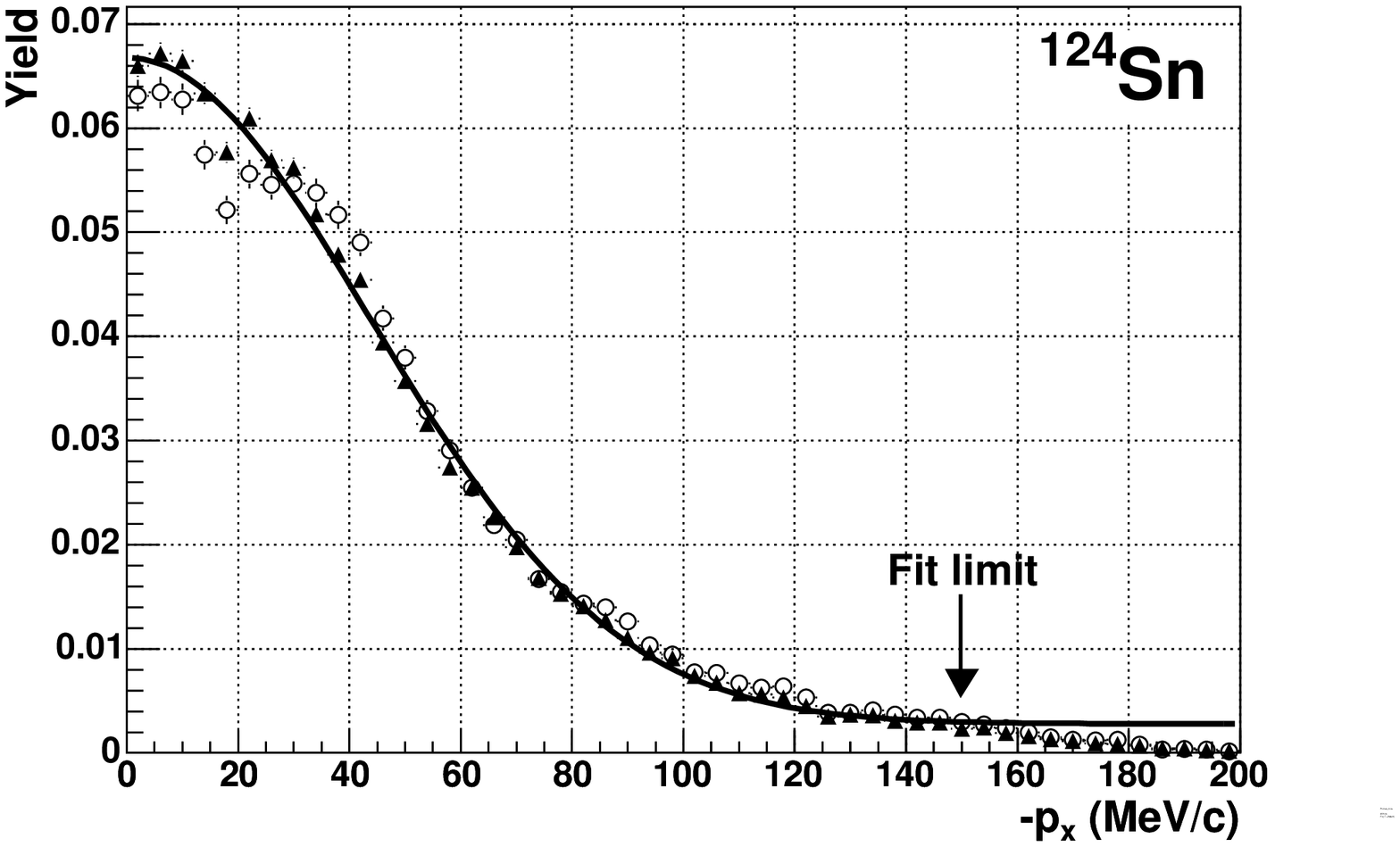}}
\caption{Distribution of transverse momentum $p_x$ for neutrons from the fragmentation 
of $^{124}$Sn projectiles in collisions with natural Sn targets at an incident energy 
of 600 MeV/nucleon, selected with the condition $Z_{\rm bound} \ge 45$.
The open 
circles and the filled triangles represent the yield of primary hits within a neutron 
cluster and the yields of all hits, respectively.
The solid line shows the result of a fit with a Gaussian distribution superimposed on a 
constant background.
}
\label{fig:px}
\end{figure} 

To determine neutron multiplicities of the spectator source , several steps had to be taken. 
The partial acceptance of LAND made it necessary to extrapolate from the kinematic regime 
covered by the detector to larger transverse momenta. This was done with Gaussian functions
in $p_x$ and $p_y$ which were found to describe well the measured spectra (Fig.~\ref{fig:px}).
Effective temperatures deduced from the distribution widths are close to $T = 4$~MeV for
$Z_{\rm bound} \le 20$ and smoothly decrease to $T \approx 2.5$~MeV at 
$Z_{\rm bound}/Z_{\rm proj} \approx 1$. No significant difference was observed between the
temperatures of the neutron-rich and neutron-poor reaction systems, consistent with the
result for isotope temperatures deduced from the yields of fragments and light charged 
particles~\cite{sfienti09}.

\begin{figure} [htb!]
\centerline{\includegraphics[width=7.7cm]{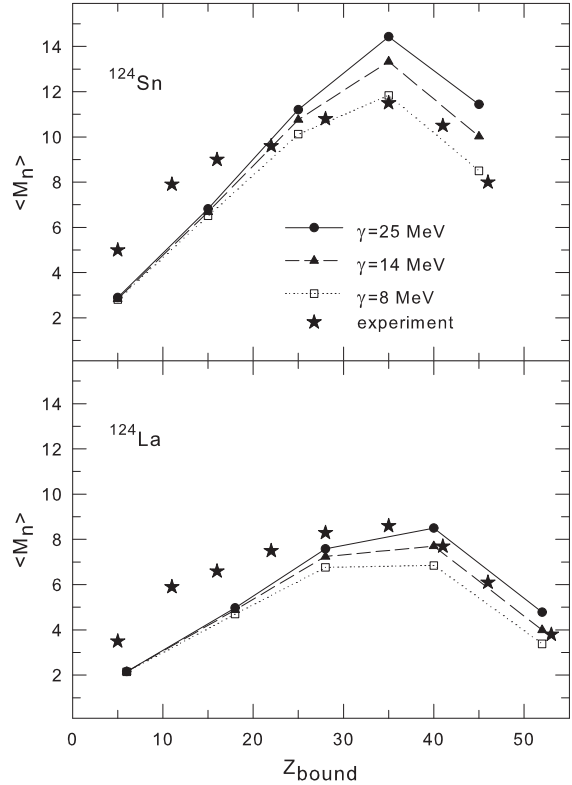}}
\caption{Mean neutron multiplicity of the spectator sources produced in the fragmentation of
$^{124}$Sn (top panel) and $^{124}$La (bottom panel) as a function of 
$Z_{\rm bound}$. The experimental data are represented by the stars. The results of the SMM 
ensemble calculations are shown for three values of the symmetry-term coefficient $\gamma$ and 
represented by the symbols and lines as indicated.
}
\label{fig:nihal1}
\end{figure}

The efficiency of LAND in this experiment for neutrons of about 600 MeV was determined with
a hit-rejection algorithm and also from the probability for zero hits obtained by fitting 
the measured hit number distributions of identified neutron events with Poissonian functions. 
Comparable results were obtained with the two methods, and a global efficiency value 
$\eta = 0.73$ was adopted. The resulting
multiplicities of the spectator neutron source are shown in Fig.~\ref{fig:nihal1}.
They clearly reflect the neutron richness of the system, exhibiting maxima of 
$M_n \approx 12$ for $^{124}$Sn and $M_n \approx 9$ for $^{124}$La at values of 
$Z_{\rm bound}/Z_{\rm proj} \approx 0.7$. They increase rapidly with increasing excitation
energy in the regime of residue production ($Z_{\rm bound}/Z_{\rm proj} > 0.8$) and drop
more smoothly from their maxima to smaller values at small $Z_{\rm bound}$.

\section{Neutrons and the symmetry energy}

Predictions for the multiplicity of neutrons from the spectator decay obtained with the
Statistical Multifragmentation Model are shown in Fig.~\ref{fig:nihal1}. They were calculated
with the ensemble and liquid-drop parameters previously determined in the study of fragment
production in these reactions~\cite{ogul11}. As expected, the neutron multiplicity is
sensitive to the symmetry term coefficient used in the calculations. A reduction of the 
symmetry term causes the neutron multiplicities to decrease, mainly however in the range of
$Z_{\rm bound}/Z_{\rm proj} > 0.5$, because more neutrons are found bound in fragments in this 
case (cf.~Fig.~\ref{fig:ogul22}). As for the fragment $\langle N \rangle/Z$, the effect
is larger for the more neutron-rich system.  

For large $Z_{\rm bound}$, the experimental neutron multiplicities are smaller than the 
predictions obtained with the standard coefficient $\gamma = 25$~MeV, consistent with the
requirement of a reduced symmetry term following from the analysis of the measured isotope 
distributions of intermediate-mass fragments. It is less evident in the case of the 
neutron-poor $^{124}$La for which the sensitivity is small. It is also clear that the
neutron multiplicities are overall more uncertain than the precisely measured fragment data.
The observed consistency is, nevertheless, rather encouraging.

The yield of neutrons in the range of $Z_{\rm bound}/Z_{\rm proj} < 0.5$ is not as well
reproduced by the present calculations, independently of the chosen coefficient $\gamma$.
Several possible reasons can account for this discrepancy. 
The assumption of an equilibrated spectator
source made in the SMM is very successful for fragments, as demonstrated repeatedly and as 
supported by the rapidity distributions measured for intermediate mass 
fragments~\cite{schuett96}. It is less evident for light particles, as shown for helium
isotopes in the same reference. In particular, in the more central collisions with smaller
$Z_{\rm bound}$, a significant tail of helium particles extends to small rapidities,
possibly representing a component of preequilibrium emissions or simply the region of
overlap with the fireball domain. This is similar for neutrons, apparently, and 
not accounted for in the calculations with parameters
adjusted to best reproduce the yields and correlations of the observed projectile fragments.

\section{Conclusions and outlook}

The emission of neutrons in the fragmentation of excited Sn and La projectiles has been studied
with the Large Neutron Detector LAND positioned forward of the ALADIN spectrometer at SIS.
The spectator source of neutrons, centered near projectile rapidity, has been identified and its
multiplicities and effective temperatures have been determined. The measured mean 
multiplicities with maximum values of 9 to 12 reflect the neutron richness of the initial 
projectile, while the temperatures ranging from 3 to 4 MeV are insensitive to it. Calculations
with the Statistical Multifragmentation Model show that the neutron multiplicity depends
on the strength of the symmetry-term coefficient used in the calculations for the liquid-drop 
description of fragments. From the study of
the fragment isotope distributions measured in the same experiment, this was expected. 
The experimental multiplicities are in better agreement with the predictions obtained with a 
reduced symmetry term and thus support the previously reached conclusion~\cite{ogul11}.
The uncertainties are large, however, because of the extrapolation and corrections needed
for the determination of absolute multiplicities. 

Besides multiplicities, also the dynamical properties of neutrons in heavy-ion collisions 
have been found useful for investigating the strength of the nuclear symmetry energy at
densities away from normal nuclear density~\cite{lipr08}. A value for the strength of the 
symmetry term at suprasaturation densities has very recently been derived from the comparison
of neutron flows with that of charged particles in Au+Au reactions at 
400~MeV/nucleon~\cite{russo11}. Further results in this direction may be expected from more 
precise neutron measurements in similar reactions.

\end{document}